\documentclass[twocolumn,preprintnumbers,superscriptaddress,prl,nofootinbib,longbibliography,amsmath,amssymb]{revtex4-1}

\usepackage{graphicx}
\usepackage{dcolumn}
\usepackage{bm}
\usepackage{color}
\usepackage{ulem}
\usepackage{gensymb}
\usepackage{braket}
\usepackage{amsmath}
\usepackage[percent]{overpic}

\begin{document}

\title{Evidence of chiral fermion edge modes through geometric engineering of thermal Hall in $\alpha$-RuCl$_3$}

\author{Heda Zhang}
\email{zhangh3@ornl.gov}
\affiliation{Materials Science and Technology Division, Oak Ridge National Laboratory, Oak Ridge, Tennessee 37831, USA}

\author{G\'abor B. Hal\'asz}
\email{halaszg@ornl.gov}
\affiliation{Materials Science and Technology Division, Oak Ridge National Laboratory, Oak Ridge, Tennessee 37831, USA}

\author{Sujoy Ghosh}
\affiliation{Center for Nanophase Materials Sciences, Oak Ridge National Laboratory, Oak Ridge, Tennessee 37831, USA}

\author{Stephen Jesse}
\affiliation{Center for Nanophase Materials Sciences, Oak Ridge National Laboratory, Oak Ridge, Tennessee 37831, USA}

\author{Thomas Z. Ward}
\affiliation{Center for Nanophase Materials Sciences, Oak Ridge National Laboratory, Oak Ridge, Tennessee 37831, USA}

\author{David Alan Tennant}
\affiliation{Department of Materials Science and Engineering, the University of Tennessee at Knoxville, Knoxville, Tennessee 37831, USA}

\author{Michael McGuire}
\affiliation{Materials Science and Technology Division, Oak Ridge National Laboratory, Oak Ridge, Tennessee 37831, USA}

\author{Jiaqiang Yan}
\affiliation{Materials Science and Technology Division, Oak Ridge National Laboratory, Oak Ridge, Tennessee 37831, USA}

\date{\today}

\begin{abstract}
The experimental observation of half-integer-quantized thermal Hall conductivity in the Kitaev candidate material $\alpha$-RuCl$_3$ has served as smoking-gun signature of non-Abelian anyons through an associated chiral Majorana edge mode. However, both the reproducibility of the quantized thermal Hall conductivity and the fundamental nature of the associated heat carriers, whether bosonic or fermionic, are subjects of ongoing and vigorous debate. In a recent theoretical work, it was proposed that varying the sample geometry through creating constrictions can distinguish between different origins of the thermal Hall effect in magnetic insulators. Here, we provide experimental evidence of chiral fermion edge modes by comparing the thermal Hall effect of a geometrically constricted $\alpha$-RuCl$_3$ sample with that of an unconstricted bulk sample. In contrast to the bulk crystals where the thermal Hall signal fades below 5\,K, the constricted crystals display a significant thermal Hall signal that remains measurable even at 2\,K. This sharp difference agrees well with the theoretical prediction and provides compelling evidence for the contribution of chiral fermion edge modes to the thermal Hall effect in $\alpha$-RuCl$_3$. More broadly, this work confirms that the geometry dependence of the thermal Hall effect can help identify chiral spin liquids in candidate materials like $\alpha$-RuCl$_3$ and paves the way for the experimental realization of thermal anyon interferometry.

\end{abstract}

\maketitle

The experimental observation of half-integer-quantized thermal Hall conductivity in $\alpha$-RuCl$_3$~\cite{kasahara2018majorana,yokoi2021half} has served as smoking-gun signature of a chiral Majorana edge mode and, hence, a non-Abelian Kitaev spin liquid~\cite{kitaev2006anyons}. Such chiral Majorana edge modes arise from the fractionalization of spin degrees of freedom into Majorana fermions---a phenomenon that has profound implications for both correlated quantum matter and fault-tolerant quantum computing~\cite{kitaev2003fault,nayak2008non}. Nevertheless, even though substantial experimental~\cite{kasahara2018unusual,kasahara2018majorana,hentrich2019large,yamashita2020sample,czajka2021oscillations,yokoi2021half,tanaka2022thermodynamics,bruin2022robustness,lefrancois2022evidence,kasahara2022quantized,bruin2022origin,czajka2023planar,lefrancois2023oscillations,zhang2023sample,zhang2024stacking,imamura2024majorana,zhang2024anisotropic,xing2025magnetothermal,leahy2017anomalous} and theoretical~\cite{ye2018quantization,vinkler2018approximately,chern2021sign,zhang2021topological,villadiego2021pseudoscalar,lee2021quantized,zhang2023spin,li2023magnonsphononsthermalhall,dhakal2025theoryintrinsicphononthermal,yang2020universal,cookmeyer2018spin,li2022thermal} efforts have been devoted to investigating the thermal transport properties of $\alpha$-RuCl$_3$, the quantization of the thermal Hall conductivity and its underlying mechanism remain the subject of intense debate. 

On the experimental front, one major source of contention is the pronounced sample dependence of thermal transport properties, primarily resulting from the stacking disorder formed during crystal growth or during the structure phase transition from the high temperature $C2/m$ to the low temperature $R\bar{3}$ space group~\cite{lee2021quantized,bruin2022origin,zhang2023sample,zhang2024stacking,zhang2024anisotropic,yamashita2020sample,kasahara2022quantized,lefrancois2022evidence,lefrancois2023oscillations,xing2025magnetothermal}. In addition, the reported thermal Hall conductivity, $\kappa_{xy}/T$, typically deviates from the exact half-integer-quantized value of $\pi k_B^2/12\hbar$, even within the putatively quantized temperature and magnetic-field range. This deviation is in stark contrast to the electrical quantum Hall effect, where the Hall resistances ($R_{xy}$) of vastly different physical systems (GaAs, graphene, etc.)~precisely conform to the same quantized value, $h/e^2$, in their respective quantized regimes. As for the origin of the thermal Hall effect in $\alpha$-RuCl$_3$, alternative explanations involving topological magnons~\cite{czajka2023planar,chern2021sign,zhang2021topological,zhang2023spin} or phonons~\cite{lefrancois2022evidence,li2023magnonsphononsthermalhall,dhakal2025theoryintrinsicphononthermal} have been proposed, challenging the Majorana-fermion scenario. These alternative explanations in terms of bosonic heat carriers are in large part motivated by the diminishing $\kappa_{xy}/T$ below $5$\,K that is consistently observed in all bulk crystals by different groups, regardless of whether a half-integer-quantized regime is present or not.

There is also an important conceptual difference between electrical transport in a quantum Hall system and thermal transport in a quantum spin liquid. In the former case, the insulating character of the bulk enables the precise observation of quantized electrical transport from the chiral edge modes. In the latter case, however, any chiral edge modes necessarily coexist with bulk heat carriers---acoustic phonons---that are coupled to the edge modes and dominate the thermal transport. Therefore, in a quantum spin liquid, the contributions of the chiral edge fermions and the bulk phonons to the thermal Hall response are extremely difficult to disentangle. In particular, while the measured $\kappa_{xy}/T$ can surprisingly remain half-integer quantized at an exceedingly small Hall angle, it is also expected to be suppressed at low temperatures due to the chiral edge fermions being thermally decoupled from the bulk phonons~\cite{ye2018quantization,vinkler2018approximately}. In other words, the diminishing $\kappa_{xy}/T$ below $5$\,K observed in bulk $\alpha$-RuCl$_3$ samples is consistent with both fermionic and bosonic origins of the thermal Hall effect.

In a recent theoretical work~\cite{theory}, it was proposed that varying the sample geometry by, for example, introducing constrictions, as shown in Figs.~1(a,b), is a viable approach for determining the origin of the thermal Hall effect. Through standard phenomenological heat-transport equations based on minimal assumptions, Ref.~\onlinecite{theory} investigated how thermal Hall effects of different origins depend on such variations in the sample geometry. It was shown that, for a conventional thermal Hall effect carried by phonons or magnons, the measured thermal Hall resistance is independent of the two-dimensional sample geometry and completely unaffected by introducing constrictions into the sample. For a thermal Hall effect resulting from a chiral fermion edge mode of a chiral spin liquid, however, the thermal Hall resistance contains an additional geometry- and temperature-dependent prefactor that is dramatically enhanced by constrictions at low temperatures. This geometry dependence of the thermal Hall resistance directly reflects the unconventional nature of the underlying thermal Hall effect.

In this Letter, we investigate the thermal Hall effect of $\alpha$-RuCl$_3$ crystals with varying geometry constrictions introduced by focused ion beam (FIB) cutting. In agreement with the theoretical prediction~\cite{theory}, a well-resolved thermal Hall signal is observed even at $2$\,K for samples with strongly constricted geometries, which is in stark contrast to the rapid suppression seen below $5$\,K in all unconstricted samples. This dramatically enhanced thermal Hall effect at low temperatures induced by geometry constrictions provides compelling evidence that chiral fermion edge modes contribute to the thermal Hall effect in $\alpha$-RuCl$_3$. More generally, this work confirms that the geometry dependence of the thermal Hall effect can help identify chiral spin liquids in candidate materials such as $\alpha$-RuCl$_3$ and paves the way for the experimental realization of thermal anyon interferometry \cite{klocke2022thermal,wei2021thermal,wei2023thermal}.

\begin{figure} \centering \includegraphics [width = 0.5\textwidth] {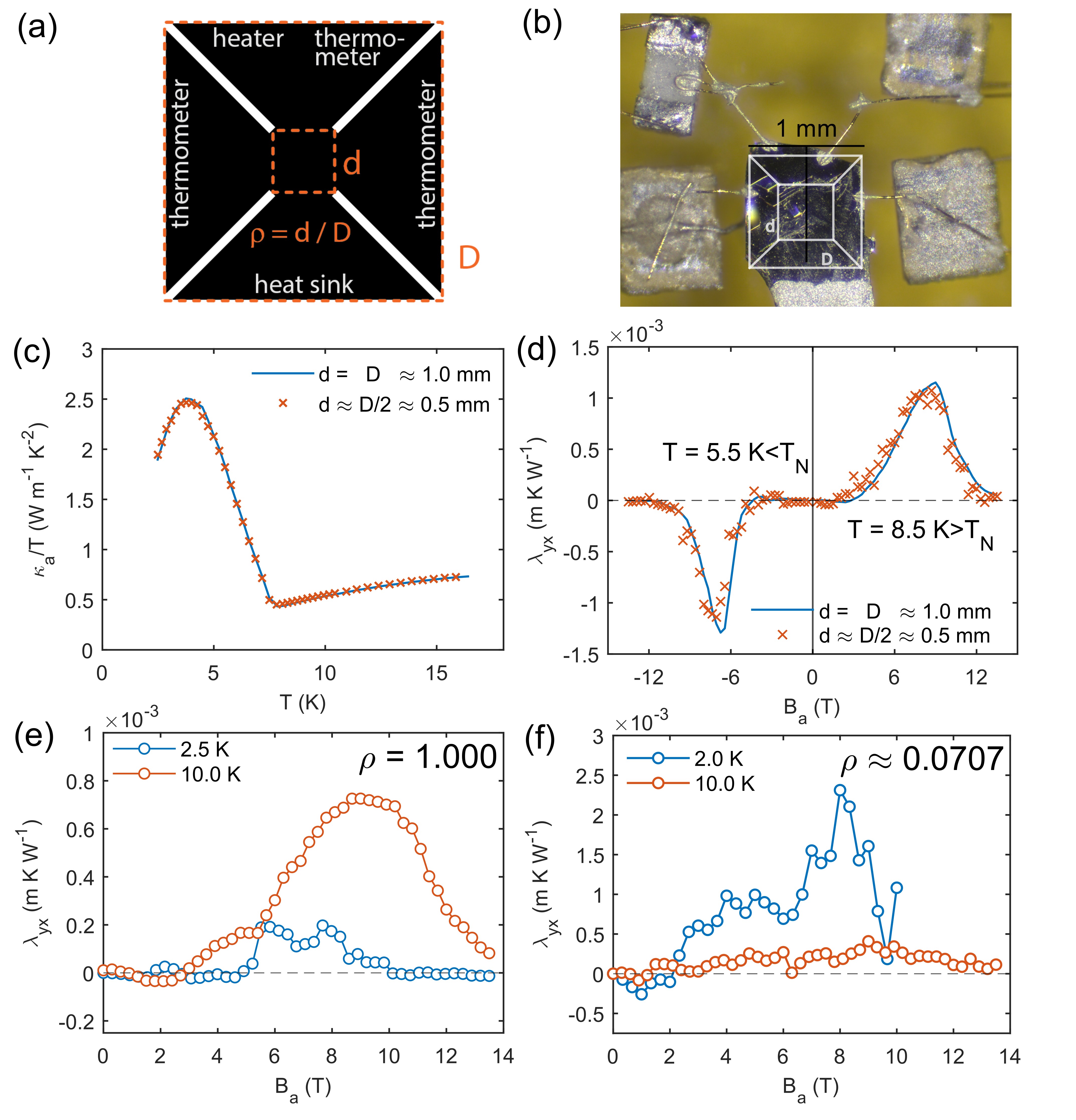}
\caption{Geometry constrictions and thermal transport properties of $\alpha$-RuCl$_3$. (a) Schematic picture of the clover-shaped four-lobe geometry distinguishing between different origins of the thermal Hall effect~\cite{theory}. A dimensionless ratio $\rho = d/D$ is introduced to describe the degree of cutting, with $\rho=1$ for a pristine crystal. (b) Picture of a crystal cut using focused ion beam (FIB) and connected with gold wires to a heater and three thermometers. (c,d) Longitudinal thermal conductivity ($\kappa_{xx}/T$) and thermal Hall resistivity ($\lambda_{yx}$) of Sample 1 before ($\rho\,=\,1$) and after ($\rho \approx 0.5$) FIB cutting. As explained in the text, FIB cutting does not degrade the $\alpha$-RuCl$_3$ crystal quality. (e,f) Field-dependent thermal Hall resistivity of Sample 1 in pristine ($\rho = 1$) and FIB-cut ($\rho \approx 0.07$) forms. In sharp contrast to the pristine $\alpha$-RuCl$_3$ crystal with $\rho=1$, the FIB-cut crystal with $\rho \approx 0.07$ has a prominent thermal Hall effect at 2\,K larger than that at 10\,K. The data shown in panels (c-f) were acquired under conditions where the magnetic field and thermal current were both perpendicular to the Ru-Ru bonds (i.e., parallel to the crystallographic $a$ axis).} 
\label{Fig-1}
\end{figure}

\begin{figure*} \centering \includegraphics [width = 0.95\textwidth] {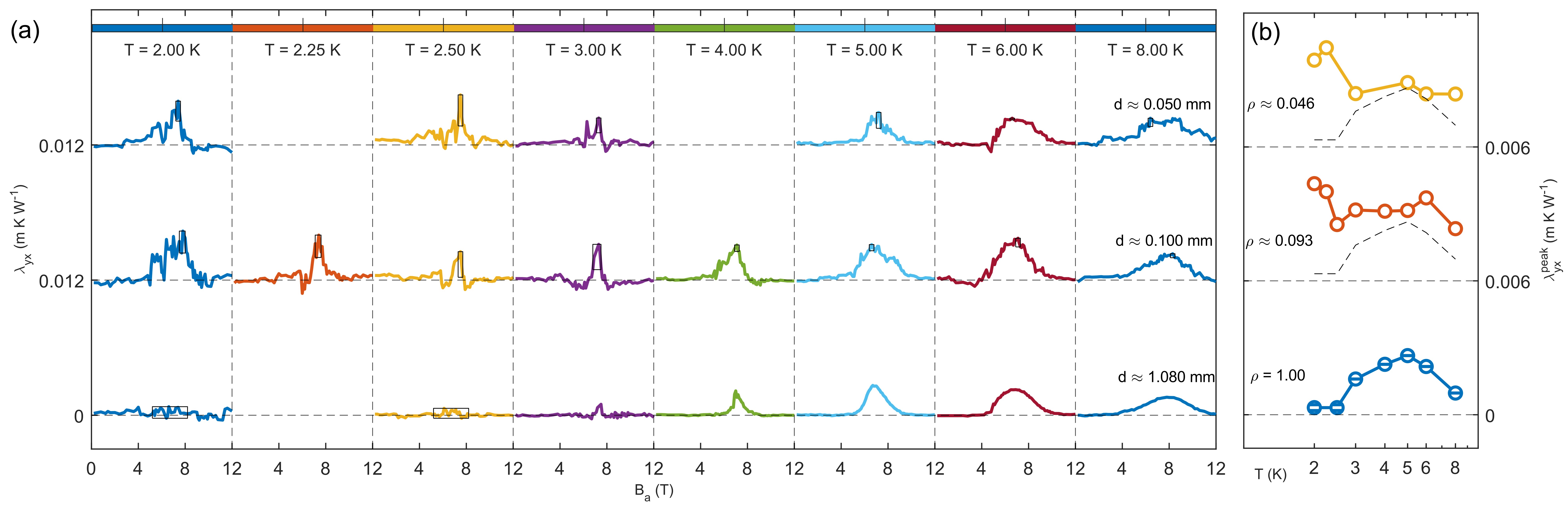}
\caption{Geometry constriction enhanced thermal Hall effect in Sample 2. (a) Field-dependent thermal Hall resistivity $\lambda_{yx}$ in the temperature range 2\,K $\leq T \leq$ 8\,K for the pristine crsytal ($\rho$\,=\,1) and the FIB-cut crystals ($\rho=0.046,0.093$). (b) The peak of $\lambda_{yx}$ for each field scan [$\lambda_{yx}^{\mathrm{peak}}(T)=\max_{B_a} \lambda_{yx}(B_a,T)$] as a function of temperature. The dashed curves are $\lambda_{yx}^{\mathrm{peak}}(T)$ data for the pristine crystal to highlight the enhanced thermal Hall effect below $5$\,K in geometrically constricted crystals. The magnetic field and heat current are both applied along the crystallographic $a$ axis (perpendicular to the Ru-Ru bonds).}
\label{Fig2}
\end{figure*}

Millimeter sized $\alpha$-RuCl$_3$ crystals with minimal amount of stacking disorder were grown by vapor transport technique~\cite{zhang2024stacking}. Magnetic and specific heat measurements confirmed these crystals show a single magnetic order at 7.6\,K. For thermal transport measurements, we first made thermal contacts from the bulk $\alpha$-RuCl$_3$ crystals to Cernox thermometers. Then the sample puck was transported to a FIB instrument to carve cuts through the sample using Xenon plasma etching. Thermal transport measurements were performed before and after each FIB processing. It is worth mentioning that this process was carried out without any additional physical contacts to either the sample or the thermometers. Figure 1(b) shows a picture of such a crystal. Samples 1 and 2 were measured in a Physical Property Measurement System (Quantum Design), while Sample 3 was measured in a Triton dilution refrigerator (Oxford Instrument). Details on the temperature profile and hysteresis effects can be found in the Supplemental Material (SM)~\cite{SM}.

We extract two key quantities from our thermal transport measurements: the longitudinal thermal conductivity $\kappa_{xx}$ and the effective thermal Hall resistivity $\lambda_{yx}$. The former one is obtained as $\kappa_{xx}=\chi P/(t\,\Delta T_x)$, where $\Delta T_x$ is a longitudinal temperature difference, $t$ is the sample thickness, $P$ is the applied power, and $\chi$ is a dimensionless geometric prefactor that generalizes the aspect ratio (i.e., length over width) from a standard rectangular geometry. This prefactor is numerically computed for each specific (pristine or FIB-cut) sample geometry and each pair of thermometers by solving the heat-transport equations in Ref.~\onlinecite{theory} assuming $\kappa_{xy}=0$. The effective thermal Hall resistivity is defined as $\lambda_{yx}=t\,\Delta T_y/P$, where $\Delta T_y$ is the field-antisymmetrized transverse temperature difference. In most thermal transport studies, this quantity is directly identified with the thermal Hall resistivity---implicitly assuming that it is an exclusive property of the underlying material that does not depend on the sample geometry. It was shown in Ref.~\onlinecite{theory}, however, that this assumption is only correct for a conventional thermal Hall effect carried by phonons or magnons. If the thermal Hall effect instead originates from a chiral fermion edge mode of a chiral spin liquid, $\lambda_{yx}$ as defined above is actually a product of the thermal Hall resistivity---a property of the material---and a dimensionless geometry-dependent prefactor. This is the reason why we introduce the term ``effective thermal Hall resistivity'' for $\lambda_{yx}$.

A natural concern when introducing geometry constrictions into $\alpha$-RuCl$_3$ using FIB is that this process may unintentionally introduce structural damage and lattice defects, e.g., stacking faults. $\alpha$-RuCl$_3$ is well known for exhibiting sample-dependent physical properties---including thermal transport properties---because of such extrinsic lattice defects. We first demonstrate the validity of our method by comparing the thermal transport properties of an $\alpha$-RuCl$_3$ crystal (Sample 1) before and after cutting with FIB. Introducing the dimensionless ratio $\rho=d/D$ [see Fig.~1(a)] to characterize the degree of cutting, with $\rho=1$ corresponding to the pristine crystal, we show in Figs.~1(c,d) that FIB cutting down to $\rho=0.5$ leads to negligible changes in both $\kappa_{xx}$ above $3$\,K and $\lambda_{yx}$ above $5$\,K~\footnote{Note that, at sufficiently high temperatures, $\lambda_{yx}$ is expected to be geometry independent even for a thermal Hall effect due to a chiral fermion edge mode~\cite{theory}.}. These results suggest that FIB cutting does not significantly degrade the crystal quality by introducing stacking disorder and can be employed to shape $\alpha$-RuCl$_3$ crystals in order to explore the impact of geometry constrictions on the thermal Hall effect.

In contrast, Figs.~1(e,f) show that the low-temperature behavior of $\lambda_{yx}$ exhibits a dramatic difference between pristine ($\rho=1$) and FIB-cut ($\rho\approx0.07$) versions of Sample 1. For the pristine crystal, $\lambda_{yx}$ is strongly suppressed below $5$\,K [see Fig.~1(e)], in agreement with all previous experimental observations \cite{czajka2023planar,zhang2023sample}. After FIB cutting to $\rho\approx0.07$, however, $\lambda_{yx}$ actually exhibits a pronounced low-temperature upturn, with $\lambda_{yx}$ at $2$\,K being significantly larger than that at $10$\,K [see Fig.~1(f)]. This result demonstrates that artificial geometry constrictions in $\alpha$-RuCl$_3$ crystals indeed lead to an enhancement of the thermal Hall effect at low temperatures, as predicted for a chiral spin liquid in Ref.~\onlinecite{theory}.

To further confirm this characteristic geometry dependence of the thermal Hall effect, we next measured $\lambda_{yx}$ of a different $\alpha$-RuCl$_3$ crystal (Sample 2) both in the pristine state ($\rho=1$) and after FIB cutting to $\rho=0.093$ and $0.046$. Figure \ref{Fig2}(a) shows the results against a magnetic field $B_a$ applied in the crystallographic $a$ direction within the temperature range $2$\,K $\leq T\leq8$\,K. At 2\,K, thermal Hall signals for $\rho=0.093$ and $0.046$ can be well resolved, similar to what is observed for Sample 1. Figure \ref{Fig2}(b) shows the peak of $\lambda_{yx}$ in the field, $\lambda_{yx}^{\mathrm{peak}}(T)=\max_{B_a} \lambda_{yx}(B_a,T)$, as a function of temperature, and further highlights the low-temperature enhancement of the thermal Hall effect in a constricted sample geometry. For the pristine crystal, $\lambda_{yx}^{\mathrm{peak}}$, as determined from the field scans in Fig.~\ref{Fig2}(a), reaches its maximum around $T=5$\,K and diminishes quickly upon further cooling. After cutting to $\rho=0.093$ and $0.046$, the temperature dependence of $\lambda_{yx}$ is qualitatively similar at $T\gtrsim5$\,K but completely different at $T\lesssim4$\,K, with $\lambda_{yx}$ persisting and even increasing down to $2$\,K. 

In the SM~\cite{SM}, we show additional results for Sample 2  cut to $\rho\approx0.02$; the data quality is significantly reduced in this case due to the limited thermal power that can be applied without overheating the sample, especially at the lowest temperatures. This observation suggests that the micron-scale constrictions required for thermal anyon interferometry~\cite{klocke2022thermal,wei2021thermal,wei2023thermal} would need alternative techniques capable of detecting much smaller temperature gradients.

\begin{figure} \centering \includegraphics [width = 0.48\textwidth] {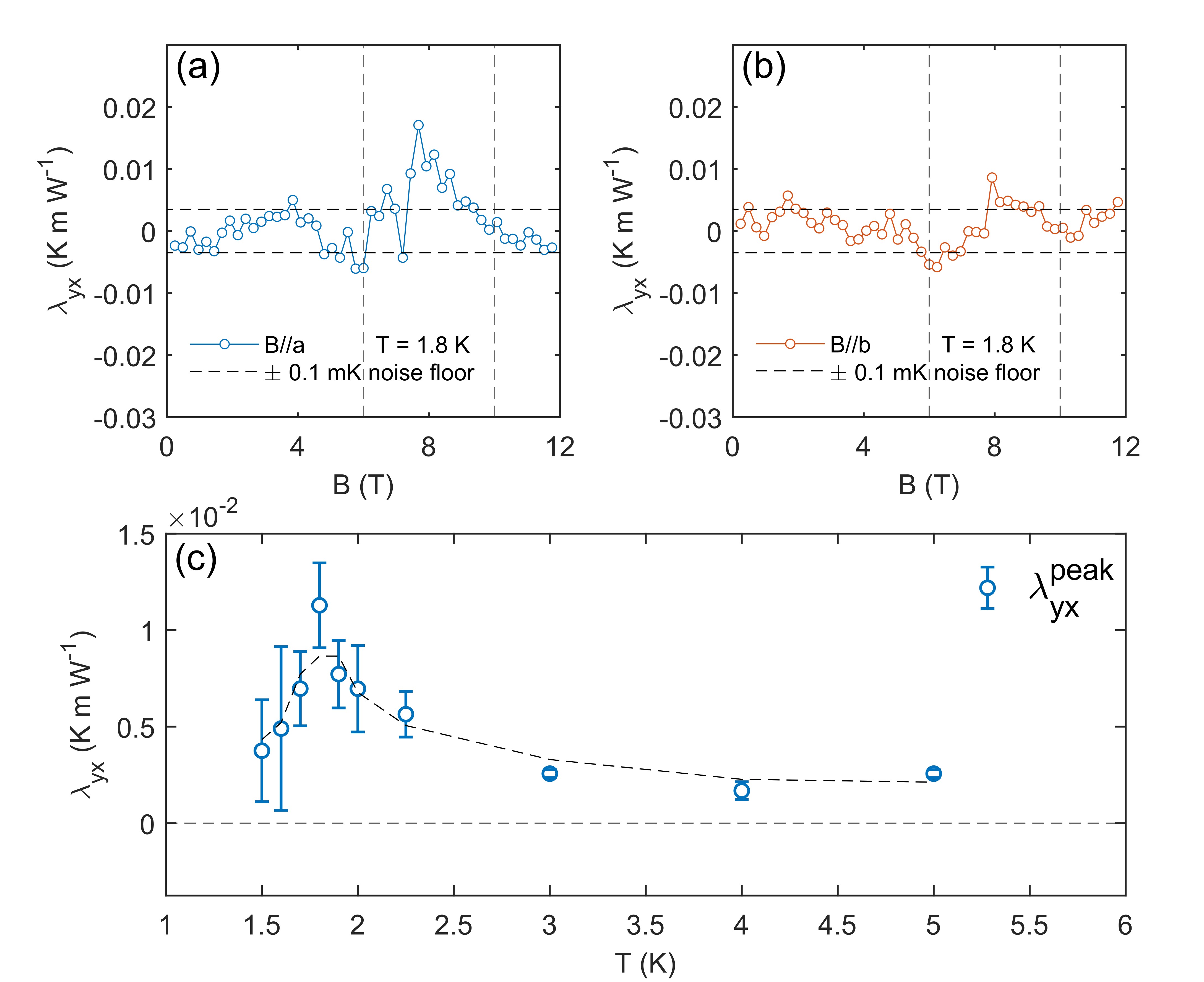}
\caption{Field-direction dependence of the geometry constriction enhanced thermal Hall effect in Sample 3 with $\rho=0.117$. (a,b) Field dependence of the thermal Hall resistivity $\lambda_{yx}$ at 1.8\,K with the magnetic field applied (a) perpendicular to the Ru-Ru bonds (in the $a$ direction) and (b) parallel to the Ru-Ru bonds (in the $b$ direction). The vertical dashed lines highlight the field range where the field-induced disordered state exists. The horizontal dashed lines indicate the noise floor set by the experimental resolution. (c) Temperature dependence of the peak thermal Hall resistivity $\lambda_{xy}^{\mathrm{peak}}$, as determined from each field scan at temperatures between 1.5\,K and 5.0\,K. The magnetic field and thermal current are both applied along the crystallographic $a$ axis. The geometry constriction enhanced thermal Hall effect is similar to that observed in Sample 2. The dashed curve is a guide to the eye. }
\label{Fig-3}
\end{figure}

To determine if the enhanced thermal Hall effect at low temperatures is an intrinsic property of $\alpha$-RuCl$_3$ with a constricted geometry, we then measured $\lambda_{yx}$ for a third crystal (Sample 3) cut to $\rho\,=\,0.117$, focusing on even lower temperatures and the field-direction dependence. Figures \ref{Fig-3}(a,b) show the results against fields $B_{a}$ and $B_{b}$ applied perpendicular and parallel to the Ru-Ru bonds, respectively, at temperature $1.8$\,K. Whereas we clearly observe a statistically nonzero $\lambda_{yx}$ around $B_a=8$\,T [see Fig.~\ref{Fig-3}(a)], the corresponding peak in $\lambda_{yx}$ around $B_b=8$\,T is comparable to the error of the measurement [see Fig.~\ref{Fig-3}(b)]. These results are consistent with an intrinsic thermal Hall effect in $\alpha$-RuCl$_3$. In particular, when the field direction coincides with the crystallographic $b$ axis, $\lambda_{yx}$ must vanish by symmetry in a $C2/m$ crystal~\cite{yokoi2021half,imamura2024majorana}, while it is expected to be significantly suppressed in an $R\bar{3}$ crystal because the same symmetry still applies within individual honeycomb layers and is only broken by layer stacking. Hence, we conclude that extrinsic factors such as experimental setups or crystal defects are unlikely to account for the enhanced thermal Hall effect at low temperatures.

Finally, Fig.~\ref{Fig-3}(c) shows the temperature dependence of $\lambda_{yx}^{\mathrm{peak}}(T)=\max_{B_a} \lambda_{yx}(B_a,T)$ for Sample 3 cut to $\rho\,=\,0.117$. The peak values of $\lambda_{yx}$ are extracted from the field scans at fixed temperatures plotted in Fig.~\ref{Fig-3}(c) and the SM~\cite{SM}. We observe that $\lambda_{yx}$ initially increases upon cooling from $5$\,K, then exhibits a peak around $1.8$\,K, and finally drops upon further cooling to $1.5$\,K. These results are consistent with the expectations for a constricted sample of a chiral spin liquid~\cite{theory}. In particular, even with constrictions giving rise to a pronounced low-temperature peak, the thermal Hall effect should eventually vanish at the lowest temperatures due to lack of thermalization between the chiral fermion edge mode and the bulk phonons~\cite{ye2018quantization}.

\begin{figure} \centering \includegraphics [width = 0.48\textwidth] {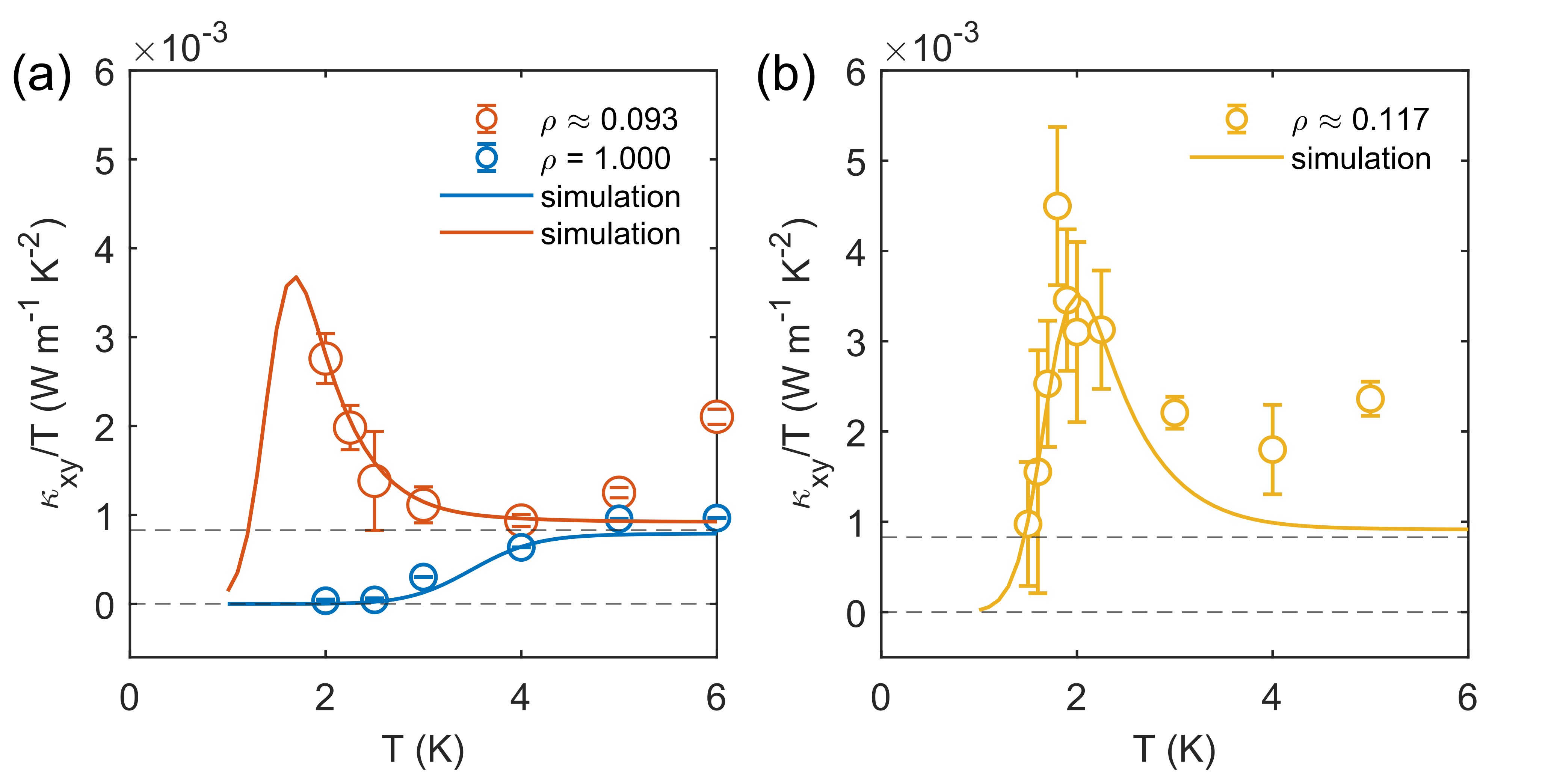}
\caption{Comparison of experimental and theoretical thermal Hall conductivities for (a) Sample 2 and (b) Sample 3. Experimental and theoretical results are shown by open circles and solid lines, respectively, displaying quantitative agreement below $5$\,K. Sample 3 was measured down to 1.5\,K in a Triton dilution refrigerator to observe the eventual suppression of the thermal Hall effect, which is expected due to reduced edge-bulk thermalization at the lowest temperatures~\cite{ye2018quantization}. The horizontal dashed lines highlight the zero and half-integer quantized thermal Hall conductivity.}
\label{Fig-4}
\end{figure}

For a more quantitative comparison with theoretical expectations, it is helpful to convert the effective thermal Hall resistivity $\lambda_{yx}$ into the corresponding thermal Hall conductivity: $\kappa_{xy}=\kappa_{xx}^2\,\lambda_{yx}$. Figure \ref{Fig-4} shows the peak of this quantity, $\kappa_{xy}^{\mathrm{peak}}=\kappa_{xx}^2\,\lambda_{yx}^{\mathrm{peak}}$, as a function of the temperature for the pristine state ($\rho=1$) of Sample 2 and the FIB-cut versions ($\rho\approx0.1$) of Samples 2 and 3. For a conventional thermal Hall effect carried by phonons or magnons, $\kappa_{xy}$ directly reduces to the thermal Hall conductivity, which is an exclusive property of the underlying material and thus independent of the sample geometry~\cite{theory}. The measured $\kappa_{xy}$ is, however, clearly enhanced at low temperatures by constrictions in the sample [see Fig.~\ref{Fig-4}(a)], which reveals that it cannot simply be a property of the material and implies that the corresponding thermal Hall effect must be of unconventional origin. In this case, $\kappa_{xy}$ is an effective thermal Hall conductivity that contains a geometry-dependent prefactor on top of the thermal Hall conductivity~\cite{theory}.

Assuming that the thermal Hall effect is connected to a non-Abelian Kitaev spin liquid~\cite{kitaev2006anyons}---the chiral spin liquid commonly hypothesized for $\alpha$-RuCl$_3$ in a field---we model the measured $\kappa_{xy}$ by considering a chiral Majorana edge mode coupled to bulk phonons~\cite{ye2018quantization}. Following Ref.~\onlinecite{theory}, we write this quantity as $\kappa_{xy}=\theta\,\kappa_{\mathrm{edge}}$, where $\theta$ is a dimensionless prefactor depending on the geometry and the temperature $T$, while $\kappa_{\mathrm{edge}} = \pi k_B^2 T/12\hbar$ (per layer) is the standard half-integer-quantized thermal Hall conductivity. In order to obtain the temperature dependence of $\kappa_{xy}$, the prefactor $\theta$ is numerically computed at each temperature by solving the heat-transport equations in Ref.~\onlinecite{theory} with $\kappa_{\mathrm{edge}} \ll \kappa_{xx}$ for the specific geometry of each pristine or FIB-cut sample. The temperature enters these equations through the edge-bulk thermalization length $\ell = \alpha T^{-5}$~\cite{ye2018quantization}, where we treat the prefactor $\alpha$ as a fitting parameter. In particular, since this prefactor is expected to depend on edge disorder~\cite{ye2018quantization}, we assume two independently fitted values for the pristine and FIB-cut edges in each sample.

The resulting theoretical curves shown in Fig.~\ref{Fig-4} provide a quantitative match to the measured $\kappa_{xy}$, capturing both the low-temperature suppression in the pristine sample and the pronounced low-temperature peak in the FIB-cut samples. The only observable discrepancies happen at higher temperatures ($T\gtrsim5$\,K) where bulk magnetic excitations may become relevant in addition to the chiral Majorana edge mode and the bulk phonons. The theoretical modelling is also useful for physically understanding the two distinct temperature regimes in the geometry dependence of $\kappa_{xy}$. While constrictions have negligible effect on $\kappa_{xy}$ for $T\gtrsim4$\,K where the thermalization length $\ell$ is small, they strongly enhance $\kappa_{xy}$ as soon as $\ell$ exceeds the constriction width $d$ at $T\lesssim3$\,K. Finally, the theoretical modelling elucidates why we see an approximately half-integer-quantized $\kappa_{xy}/T$ at $3$\,K $\lesssim T \lesssim 5$\,K in the FIB-cut version but not the pristine version of Sample 2 [see Fig.~\ref{Fig-4}(a)]. In addition to constricting the sample geometry, FIB cutting creates more disordered edges that facilitate edge-bulk thermalization~\cite{ye2018quantization}, thereby extending the range of quantization to lower temperatures.

In summary, we have utilized the thermal Hall effect in $\alpha$-RuCl$_3$ crystals with constricted geometries to detect the chiral Majorana edge mode associated with a potential non-Abelian Kitaev spin liquid. Consistent with theoretical predictions~\cite{theory}, geometry constrictions lead to an enhanced thermal Hall signal below $5$\,K that remains clearly resolvable down to $2$\,K. This observation is inconsistent with a purely bosonic origin of the thermal Hall effect in $\alpha$-RuCl$_3$ and strongly supports the existence of chiral fermion edge modes.

Taking a broader perspective, the geometrically constricted samples in this work can be viewed as prototype devices that are useful for identifying quantum spin liquids in candidate materials and may also have major technological impact. Pushing down the constriction width to the micrometer scale, analogous devices~\cite{klocke2022thermal,wei2021thermal,wei2023thermal} could provide ``smoking-gun'' evidence of a non-Abelian quantum spin liquid in $\alpha$-RuCl$_3$ and directly demonstrate non-Abelian braiding statistics via thermal anyon interferometry---a major step toward topological quantum computing~\cite{nayak2008non,kitaev2003fault}. At the same time, this work shows clear evidence of edge-mediated, field-reversible energy transfer between different parts of a millimeter-size sample and thus highlights the viability of magnetic chiral edge modes for long-distance spin and energy transmission in future spintronic devices.

\section{Acknowledgment}
JY thanks Takashi Kurumaji and Feng Ye for helpful discussions. This work was supported by the U.S. Department of Energy, Office of Science, National Quantum Information Science Research Centers, Quantum Science Center. TW was supported by the U.S. Department of Energy, Office of Science, Basic Energy Sciences, Materials Sciences and Engineering Division. FIB cutting was conducted as part of a user project at the Center for Nanophase Materials Sciences (CNMS), which is a US Department of Energy, Office of Science User Facility at Oak Ridge National Laboratory.

This manuscript has been authored by UT-Battelle, LLC, under Contract No. DE-AC0500OR22725 with the U.S. Department of Energy. The United States Government retains and the publisher, by accepting the article for publication, acknowledges that the United States Government retains a non-exclusive, paid-up, irrevocable, world-wide license to publish or reproduce the published form of this manuscript, or allow others to do so, for the United States Government purposes. The Department of Energy will provide public access to these results of federally sponsored research in accordance with the DOE Public Access Plan (http://energy.gov/downloads/doe-public-access-plan).

\section{References}
%

\end{document}